\documentclass[
 reprint,
superscriptaddress,
 amsmath,amssymb,
 aps,
 twocols,nofootinbib
]{revtex4-2}
\usepackage{verbatim}
\usepackage{graphicx}
\usepackage{dcolumn}
\usepackage{bm}
\usepackage{mathtools}  
\usepackage{booktabs}                              
           
\usepackage{float}                                    
      
\usepackage{caption}                                 
\usepackage{subcaption}                               
\usepackage{hyperref}        
\usepackage{multirow}
\usepackage{hhline}                         
\usepackage[usenames, dvipsnames]{color}
\usepackage{color}
\usepackage{xr}
\usepackage{cases}
\usepackage[normalem]{ulem,xcolor}

\newcolumntype{P}[1]{>{\centering\arraybackslash}p{#1}}

\begin{document}

\preprint{APS/123-QED}

\title{A multi-scale analysis of the CzrA transcription repressor\\ highlights the allosteric changes induced by metal ion binding}

\author{Marta Rigoli}
\affiliation{Computational mOdelling of NanosCalE and bioPhysical sysTems, Istituto Italiano di Tecnologia, Via Enrico Melen, 83 I-16152 Genoa, Italy}
\affiliation{CIBIO Department, University of Trento, Via Sommarive, 9 I-38123 Trento, Italy}
\author{Raffaello Potestio}
\affiliation{Physics Department, University of Trento, Via Sommarive, 14 I-38123 Trento, Italy}
\affiliation{Trento Institute for Fundamental Physics and Applications -- INFN TIFPA, Via Sommarive, 14 I-38123 Trento, Italy}
\email{raffaello.potestio@unitn.it}
\author{Roberto Menichetti}
\affiliation{Physics Department, University of Trento, Via Sommarive, 14 I-38123 Trento, Italy}
\affiliation{Trento Institute for Fundamental Physics and Applications -- INFN TIFPA, Via Sommarive, 14 I-38123 Trento, Italy}

\date{\today}

\begin{abstract}
Allosteric regulation is a widespread strategy employed by several proteins to transduce chemical signals and perform biological functions. Metal sensor proteins are exemplary in this respect, e.g., in that they selectively bind and unbind DNA depending on the state of a distal ion coordination site. In this work, we carry out an investigation of the structural and mechanical properties of the CzrA transcription repressor through the analysis of microsecond-long molecular dynamics (MD) trajectories; the latter are processed through the mapping entropy optimisation workflow (MEOW), a recently developed information-theoretical method that highlights, in an unsupervised manner, residues of particular mechanical, functional, and biological importance. This approach allows us to unveil how differences in the properties of the molecule are controlled by the state of the zinc coordination site, with particular attention to the DNA binding region. These changes correlate with a redistribution of the conformational variability of the residues throughout the molecule, in spite of an overall consistency of its architecture in the two (ion-bound and free) coordination states. The results of this work corroborate previous studies, provide novel insight into the fine details of the mechanics of CzrA, and showcase the MEOW approach as a novel instrument for the study of allosteric regulation and other processes in proteins through the analysis of plain MD simulations.
\end{abstract}

\maketitle

\section{Introduction}

The ArsR/SmtB family of prokaryotic metal sensor proteins is the largest and most functionally diverse metalloregulatory protein group, making it also the most extensively investigated one \cite{busenlehner2003smtb,giedroc2007metal,reyes2011metalloregulatory,antonucci2017arsr,roy2018silico,qiu2023molecular}. It includes molecules that bind a large variety of metal ions, such as Zn(II), Ni(II), and Co(II), as well as As(III), Cd(II) and Pb(II). ArsR/SmtB transcription factors regulate genes that are responsible for detoxifying the cytosol from metal ions in excess. When they bind to DNA, they repress the transcription of downstream genes, while the coordination with metals is responsible for transcriptional derepression as it induces a transition to a low-affinity state of the protein, which in turn results in the dissociation from DNA \cite{giedroc2007metal,ma2009coordination,capdevila}. Their behaviour as mechanical switches operated through metal ion binding thus makes them key elements of the cell maintenance apparatus \cite{osman2010bacterial,antonucci2017arsr,saha2017metal}, as well as a commonly employed model to understand metal-operated allosteric regulatory pathways \cite{Chemrev2016_nussinov}.

\begin{figure}[h!]
\centering
\includegraphics[width=\columnwidth]{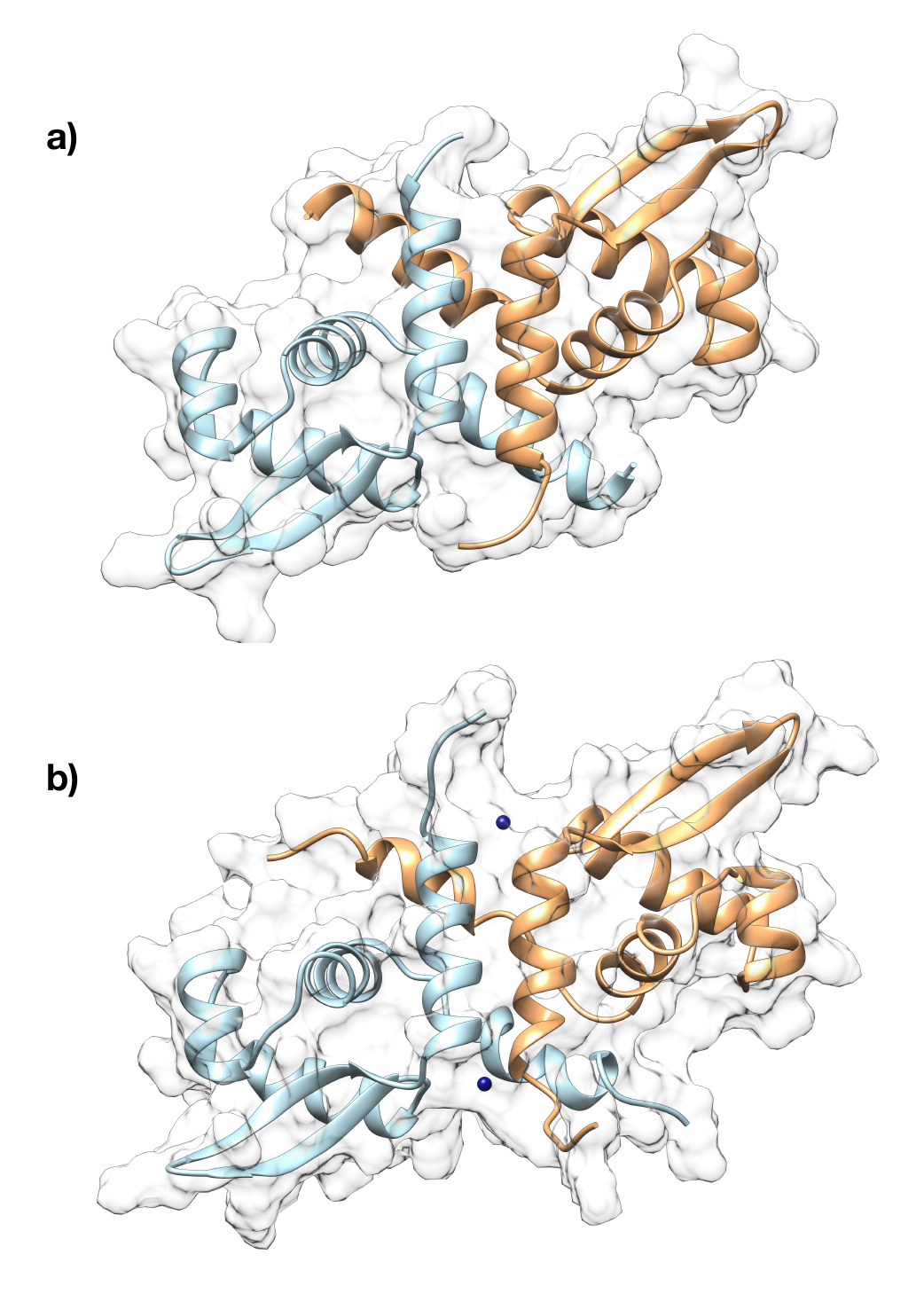}
\caption{a) Apo structure of CzrA. Chain A is shown in light blue, and chain B in orange; b) holo structure of CzrA. Chain A is shown in light blue, chain B in orange, and Zn ions are depicted in dark blue.}
\label{fig:czra_holo}
\end{figure}

In the ArsR/SmtB group one finds CzrA, a transcription repressor regulated by the concentration of Zn(II) and Co(II), which has been used as testbed for the investigation of zinc homeostasis in cell biology \cite{Chakravorty,ma2009coordination,baksh}. An allosteric mechanism is thought to be responsible for the lower affinity of the CzrA dimer for the DNA CzrO operator upon Zn(II) ion binding \cite{Chakravorty,capdevila,baksh}, this hypothesis being supported by structural studies with both experimental (e.g. NMR \cite{arunkumar2009solution,capdevila}) and computational \cite{Chakravorty} techniques. 
In these works, it was demonstrated that zinc coordination remodulates the internal dynamics with respect to the apo state, lowering the protein's binding affinity towards the DNA substrate while largely preserving its overall architecture; in contrast, a sensible structural rearrangement with respect to the apo form is visible when the molecule is in complex with the DNA substrate, in which case the protein is found in a ``bent'' conformation \cite{arunkumar2009solution,Chakravorty,capdevila} Fig. \ref{fig:czra_holo} illustrates the apo and holo (i.e. zinc-bound) forms of CzrA.

The aforementioned studies described in detail the structural organisation of CzrA and provided a convincing picture of the local and global changes that occur upon binding with the DNA and/or the zinc ions; in particular, the computational investigation carried out by Chakravorty and coworkers \cite{Chakravorty} has demonstrated that relevant insight can be gained from molecular dynamics (MD) simulations of this molecule. In the present work, we conducted an \textit{in silico} study of CzrA which builds upon these results and furthers them, with a focus on the differences that can be observed between the apo and the zinc-bound states of the protein in the absence of the DNA substrate.
More specifically, microsecond-long, all-atom molecular dynamics simulations of CzrA in the apo and holo (zinc-bound) conformations were here perfomed and analysed through the mapping entropy optimisation workflow (MEOW) \cite{giulini,PRE2022_holtzman,giulini2021system,aldrigo2024low}; the latter is an information-theoretic method recently developed by some of us---and freely available within the EXCOGITO software suite \cite{giulini2024excogito}---that aims at highlighting subsets of atoms that play a key role in the collective behaviour of a molecular system.
Following MEOW, CzrA was inspected through the lenses of its maximally-informative reduced representations, namely low-resolution, or coarse-grained, descriptions of the protein that, despite a reduction in the number of degrees of freedom employed to \emph{observe} the system, are able to retain the largest amount of statistical information on the original, all-atom reference. We underline that the theoretical foundations of MEOW are such that
the mechanical and functional insight that emerges about the protein is intrinsically multi-body---i.e. it is not straightforwardly decomposable in terms of the interplay among few constituent atoms \cite{dijkstra1999phase,likos2001effective,d2015coarse,menichetti2017thermodynamics}.

The application of MEOW to the MD simulations of the apo and holo states of CzrA allowed us to highlight the effect of metal coordination on the molecular structure, with the change in the binding state of the zinc pockets reverberating on an alteration of the properties of the whole protein, including the distal DNA binding region.
Interestingly, this alteration is not accompanied by a major rearrangement of the structure, which, as anticipated, only takes place when the protein is bound to the DNA substrate alone \cite{Chakravorty,capdevila,baksh}. The MEOW analysis is thus shown to be capable of pinpointing key features of the protein's allosteric regulation that occurs upon zinc coordination. These results, which corroborate the current understanding of this molecule's functioning built by previous studies and extend their scope, are instrumental in establishing the MEOW analysis protocol as a novel and enabling tool that can contribute to the investigation of allosteric proteins, perspectively also in the absence of previous information on the substrates and/or binding regions of the system of interest.

In Sec.~\ref{sec:res_disc}, we report the results of the MD simulations of CzrA and their subsequent investigation, while in Sec.~\ref{sec:conclusions} we sum up their implications and provide our concluding remarks. A detailed description of the system setup and the tools employed for the production and analyses of the data are provided in Sec.~\ref{sec:methods}.

\section{Results and discussion}
\label{sec:res_disc}

\subsection{Structural and dynamical analysis of the trajectories}
\label{sec:structural_analysis}

We performed $1~\mu$s all-atom molecular dynamics simulations of the CzrA transcription repressor in both the apo (PDB code 1R1U) and holo (PDB code 2M30) states of its zinc coordination site, see Sec.~\ref{sec:methods_allatom} for the associated technical details. To set the stage for the subsequent investigation of the two systems carried out through the MEOW protocol, we will first discuss the outcome of a characterisation of the MD trajectories in terms of their ensemble structural properties, in line with the study performed by Chakravorty and coworkers in Ref.~\cite{Chakravorty}.

\begin{figure}[h!]
\centering
\includegraphics[width=\columnwidth]{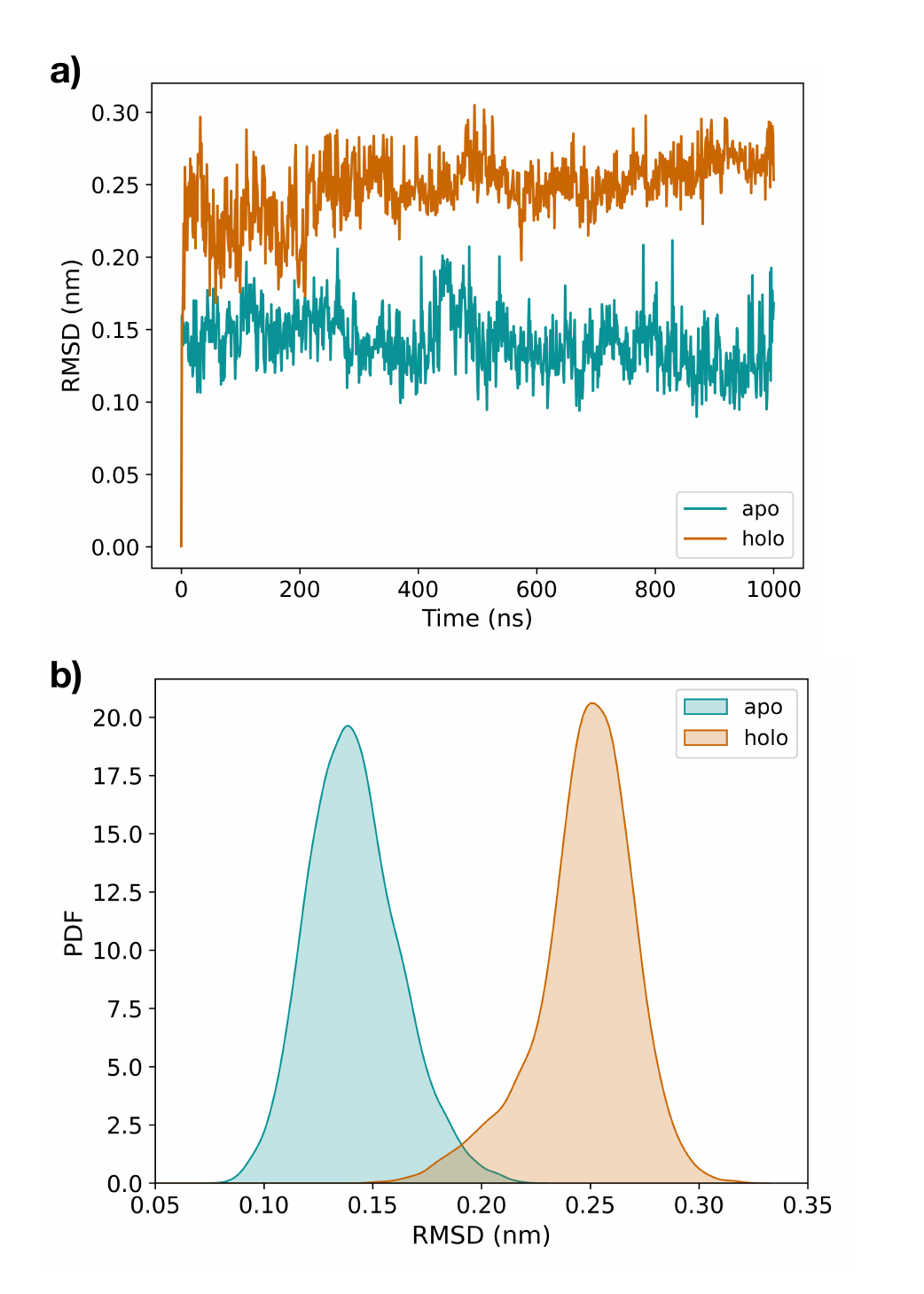}
\caption{a) Time series of the root mean square deviation (RMSD) values of the C$_{\alpha}$ atoms of the apo (light blue) and holo (orange) forms of CzrA, calculated with respect to the energy-minimised, experimental PDB configurations of the two systems; b) distributions of the time series in panel a).}
\label{fig:rmsd}
\end{figure}

For each analysed form of CzrA, in Fig.~\ref{fig:rmsd} we display the time series of the root mean square deviation (RMSD) of its C$_{\alpha}$ atoms calculated with respect to the energy-minimised PDB configuration (panel a) as well as the corresponding distribution (panel b). A qualitative inspection of these plots suggests that, within the investigated timescale of $1~\mu$s, both the apo and holo systems individually explore a single conformational basin---in principle different for each state of the protein---with fairly comparable global fluctuations; this can be deduced by the fact that the two distributions in Fig.~\ref{fig:rmsd}b appear to be unimodal and characterised by similar variances. At the same time, the apo form presents a lower average value of RMSD (namely $\sim$ 0.15 nm) compared to the one of the holo form ($\sim$ 0.25 nm); this is indicative of the 1R1U crystallographic structure of CzrA being closer to its ``equilibrium'' conformation than the zinc-bound 2M30 one, albeit by a relatively small amount.

The heuristic picture for the variability of the apo form of CzrA in terms of conformational basins that emerges from our analysis of its RMSD differs from the corresponding one reported in the work of Chakravorty \emph{et al.} \cite{Chakravorty}, where we make particular reference to their $120$~ns simulations performed \emph{in the absence} of the DNA substrate. Indeed, in Ref.~\cite{Chakravorty} the RMSD distribution of the zinc-bound protein was found to be unimodal as in our Fig.~\ref{fig:rmsd}b, while the apo system was observed to explore three distinct states, with two lateral, metastable free energy minima being separated from a central stable one by relatively low barriers. One of the aforementioned metastable basins was further linked to the stable conformation sampled by their simulations of the holo form, thus suggesting the existence of this latter state also in the configurational space accessible to the apo structure. As previously stated, our results instead qualitatively hint at the presence of a \emph{single} basin associated with each system; it is thus logical to assess the degree of similarity between these two conformational states, investigating to what extent the coordination with the metal ions results in a readjustment of what, within the analysed timescale, appears to be the one equilibrium conformation of apo-CzrA. To this aim, we performed an RMSD-based clustering of the two trajectories by relying on the UPGMA algorithm \cite{sokal1958statistical}, arbitrarily fixing the distance threshold employed in the analysis of each simulation such that the associated configurations were partitioned in no more than $10$ distinguishable macrostates. This resulted in the apo trajectory being divided into $9$ clusters, while the holo one was broken down to $7$. The most representative cluster for each form of CzrA was then identified, comprising $\sim89\%$ and $\sim55\%$ of the total number of apo and holo MD frames, respectively, and its central structure extracted; finally, the RMSD between the C$_{\alpha}$ atoms of these two configurations was calculated, amounting at $\sim 0.11$~nm and with the largest discrepancies between the two structures being found on the protein termini. Overall, these results suggest that, in the absence of the DNA substrate and on the timescale explored, the global molecular \emph{architecture} of CzrA is only marginally affected by the binding with the zinc ions, with the protein fluctuating around an average conformation that is quite compatible between the apo and holo forms \cite{arunkumar2009solution,capdevila}.

\begin{figure}[h!]
\centering
\includegraphics[width=\columnwidth]{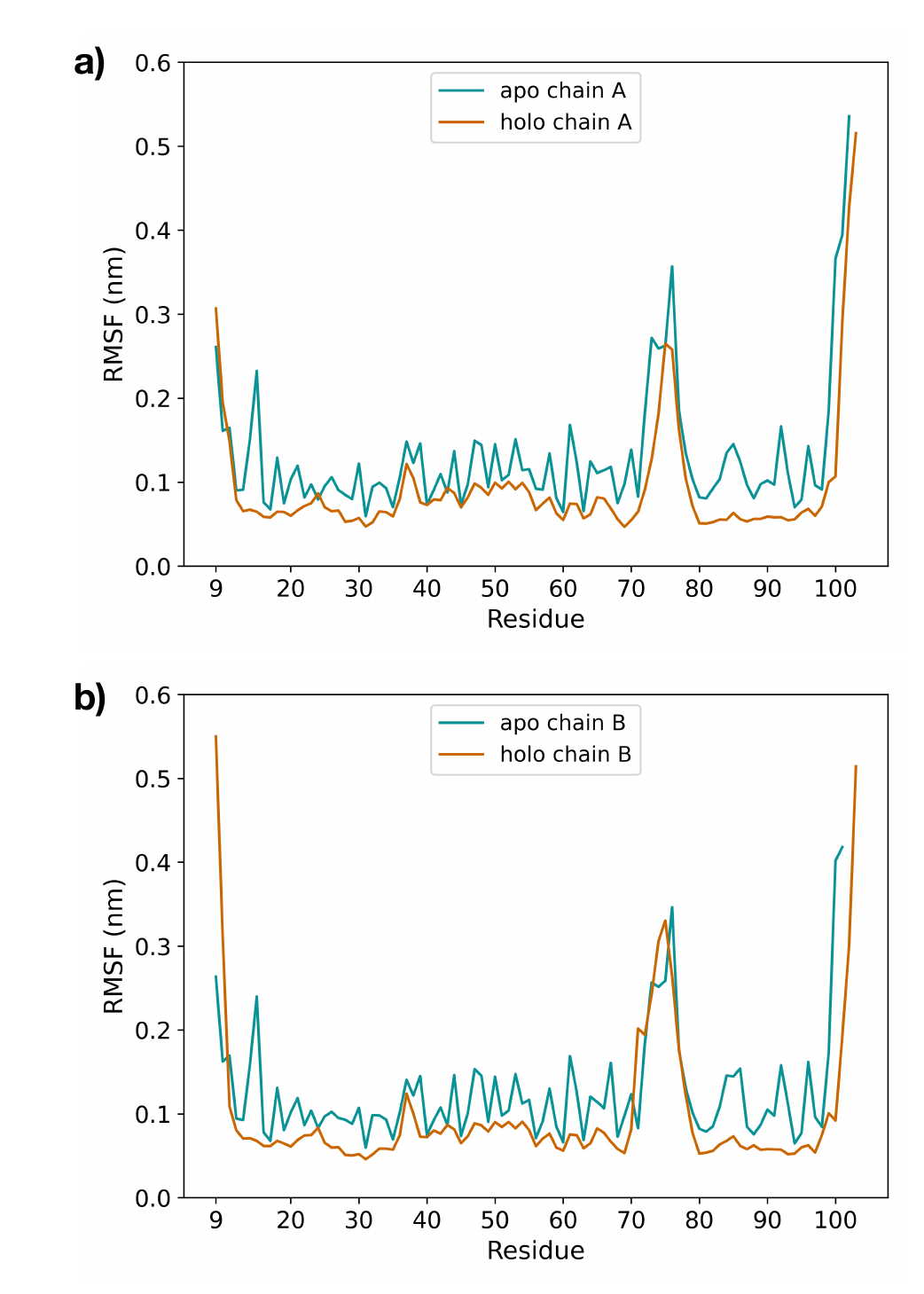}
\caption{Root mean square fluctuations (RMSF) of the C$_\alpha$ atoms of CzrA for the apo and holo forms, reported separately for each chain. a) Results for chain A of the apo (light blue) and holo (orange) systems; b) results for chain B of the apo (light blue) and holo (orange) systems.}
\label{fig:rmsf}
\end{figure}

The next natural step in the analysis is to investigate how the zinc coordination state impacts the protein's fluctuation \emph{patterns} around such conformations. The structural variability of the two forms of CzrA is thus inspected on a residue basis in Fig.~\ref{fig:rmsf}, where we compare the apo and holo systems' root mean square fluctuations (RMSF) of the C$_\alpha$ atoms calculated over all the MD trajectory frames. The comparison is presented separately for the two chains of the protein, in that slight discrepancies can be observed between their RMSF in both systems; these variations can be ascribed to a structural asymmetry of the two identical monomers composing the molecule, an asymmetry that is already present in the 1R1U and 2M30 PDB configurations and that persists throughout the associated MD simulations. A general trend can nonetheless be appreciated when comparing the two structures, namely that, consistently with what was observed in previous computational and experimental studies \cite{Chakravorty,arunkumar2009solution}, the holo system typically features lower values of RMSF compared to the apo one. This reduction in mobility is particularly evident in the stretch between residues $80$ and $100$ of each monomer---that is, in the region containing the binding sites of the zinc ions---and globally suggests that the molecule displays an increased rigidity upon coordination with the metal. The only significant exception to this feature of the holo state is given by the $\beta$ wing regions located approximately between residues $70$ and $80$ of both chains, which wave comparably to how they do in the apo form \cite{Chakravorty,arunkumar2009solution}.

Another crucial aspect regarding the configurational variability of the two systems is their capability to explore ``open'' and ``closed'' conformations, respectively characterised by low and high binding affinities with the DNA substrate \cite{arunkumar2009solution}. A simple yet useful metric to discriminate such states is the inter-protomer distance between the Ser54 serine residues of each chain \cite{Chakravorty,Chakravorty1}; relying on this collective coordinate, in their computational study \cite{Chakravorty} Chakravorty and coworkers classified as \textit{open} those configurations of their MD simulations in which the distance between the two serines was larger than $4.3$~nm, and \textit{closed} those displaying a distance below $4.1$~nm. Along their two $120$~ns trajectories of apo- and holo-CzrA in the absence of the DNA substrate, both systems were found to sample, on average, an open conformation, and no marked discrepancy was observed between the patterns of their inter-protomer distances, again suggesting the overall structural similarity of the two forms. At the same time, the zinc-bound protein was detected to seldom explore a ``flat'' conformation characterised by a reduced DNA binding affinity, displaying a serine-serine distance of $\sim 4.8$~nm. On the other hand, in their simulation of the apo form they did not observe the closed structure of the molecule that occurs in the quaternary CzrA$\bullet$DNA complex, whose serine-serine distance, according to the 2KJB ensemble of NMR structures \cite{arunkumar2009solution}, amounts instead to $\sim 4.0$~nm. We underline that, in Ref.~\cite{Chakravorty}, closed conformations akin to the experimental ones only appeared in MD simulations of apo-CzrA that were either (\emph{i}) initiated from the 2KJB structure to investigate the protein's closed-to-open transition; or (\emph{ii}) performed in the presence of a DNA substrate explicitly bound to the molecule.

\begin{figure}[]
\centering
\includegraphics[width=\columnwidth]{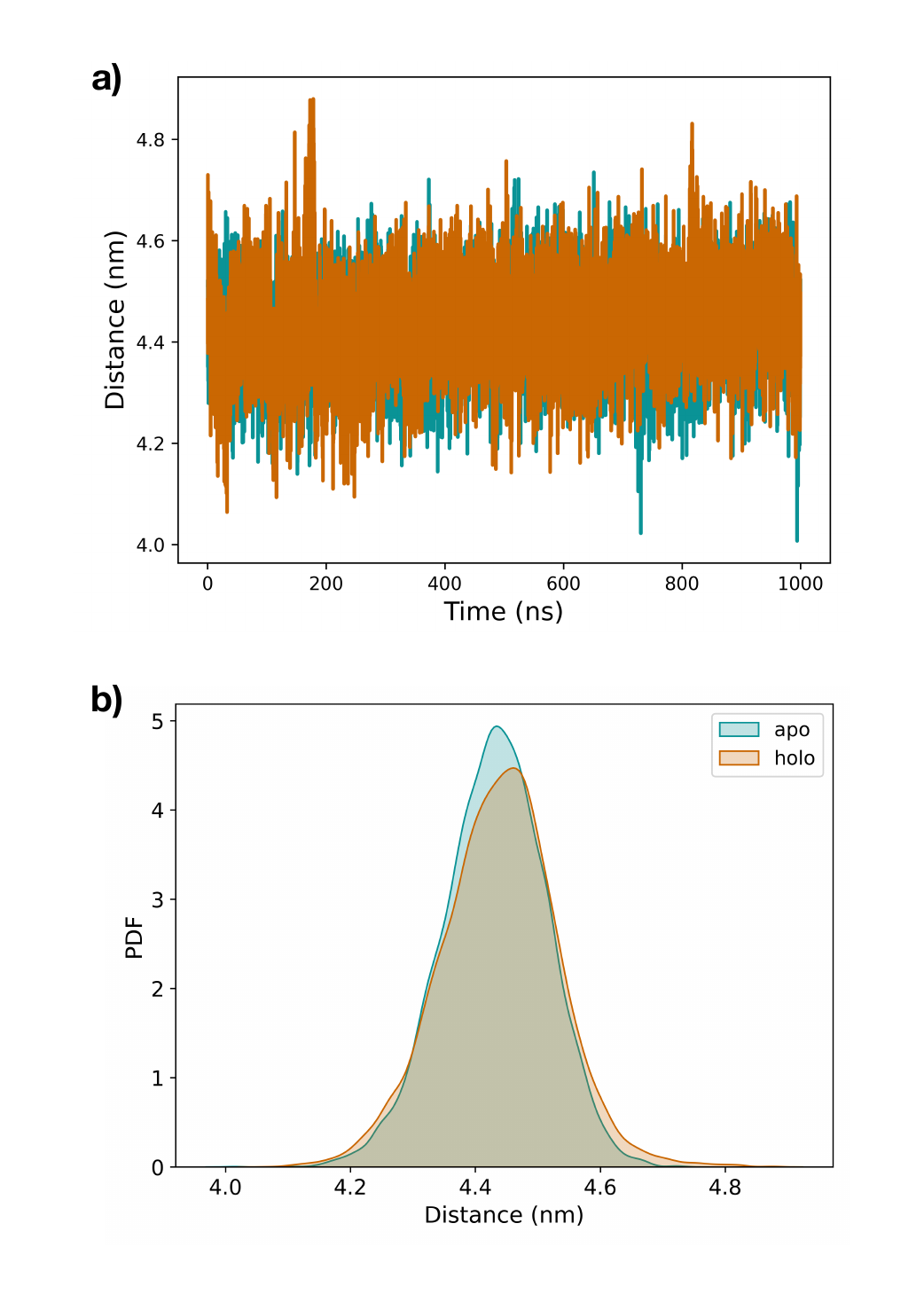}
\caption{a) Time series of the distance between the C$_\alpha$ atoms of Ser54.A and Ser54.B \cite{Chakravorty}, for both the apo (light blue) and holo (orange) forms of CzrA; b) distributions of the time series in panel a).}
\label{fig:distance}
\end{figure}

Consistently with these analyses, we measured the distance between the two serines in each frame of our MD trajectories for both the apo and holo forms of CzrA. From the time series of the inter-protomer distances presented in Fig.~\ref{fig:distance}a, it is possible to appreciate that the behaviour of the two systems is largely compatible; the associated distributions, reported in Fig.~\ref{fig:distance}b, indeed exhibit very similar variabilities and average values, where all configurations are characterised by serine distances lying in the range between $4$ and $5$~nm, and a mean distance of $\sim4.4$~nm is observed in both forms. These trends are in line with the ones reported in the work of Chakravorty \emph{et al.} \cite{Chakravorty}, again referencing their MD simulations of apo- and holo-CzrA performed in the absence of DNA. This notwithstanding, the data for the two forms also feature small but noticeable discrepancies, in particular the presence of spikes in the time series that, in the case of the holo system, reach values up to $\sim 4.9$ nm, with the molecule adopting a ``flat'', less DNA-binding prone conformation---already observed in Ref.~\cite{Chakravorty}---due to the coordination with the zinc ions. As for the apo state, we interestingly note that within our analysed timescale the serine-serine distance attains instead values as low as $\sim 4.0$ nm, hence in agreement with the experimental NMR structures of the closed conformation of CzrA in complex with DNA \cite{arunkumar2009solution}. This suggests that the protein in the apo state is capable of making excursions to a state of high DNA binding affinity even in the absence of the latter; whether this is sufficient evidence of a binding mechanism uniquely or mainly relying on conformational selection is hard to infer, since the fraction of simulated time spent in the lowest inter-serine distance configuration is rather small, see Fig. \ref{fig:distance}. At this stage, the hypothesis of induced fit seems to be more substantiated.

Summarizing the results gathered insofar, our MD simulations show that the apo and holo forms of the protein are rather similar in terms of overall architecture of the molecule; in contrast, appreciable differences can be observed in their \emph{dynamics}, with zinc coordination significantly dampening the conformational variability of CzrA around the equilibrium configuration characteristic of the investigated timescale, as highlighted by the fluctuation patterns of the two structures presented in Fig.~\ref{fig:rmsf}.

\begin{figure}[]
\centering
\includegraphics[width=\columnwidth]{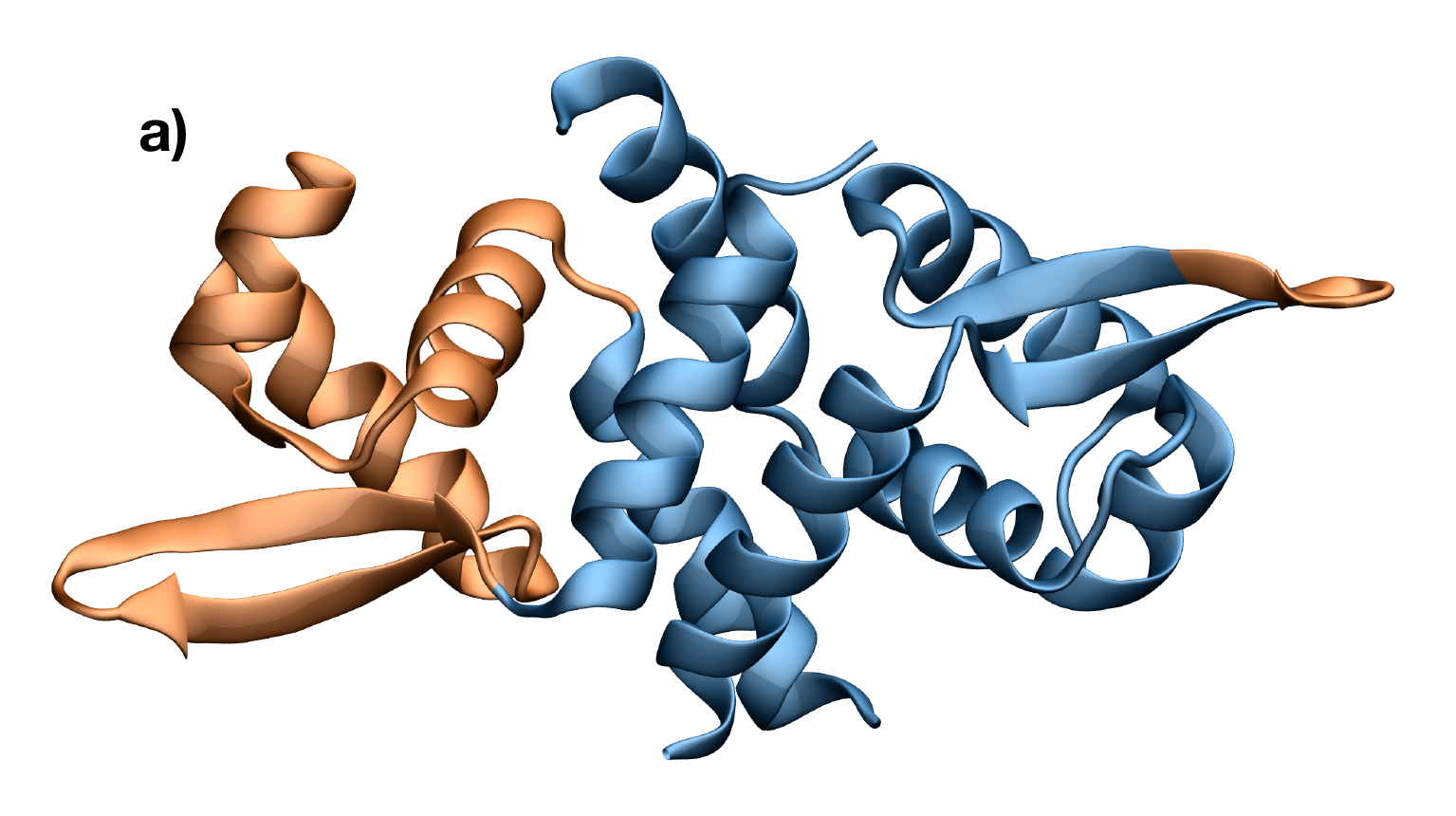}\\
\includegraphics[width=\columnwidth]{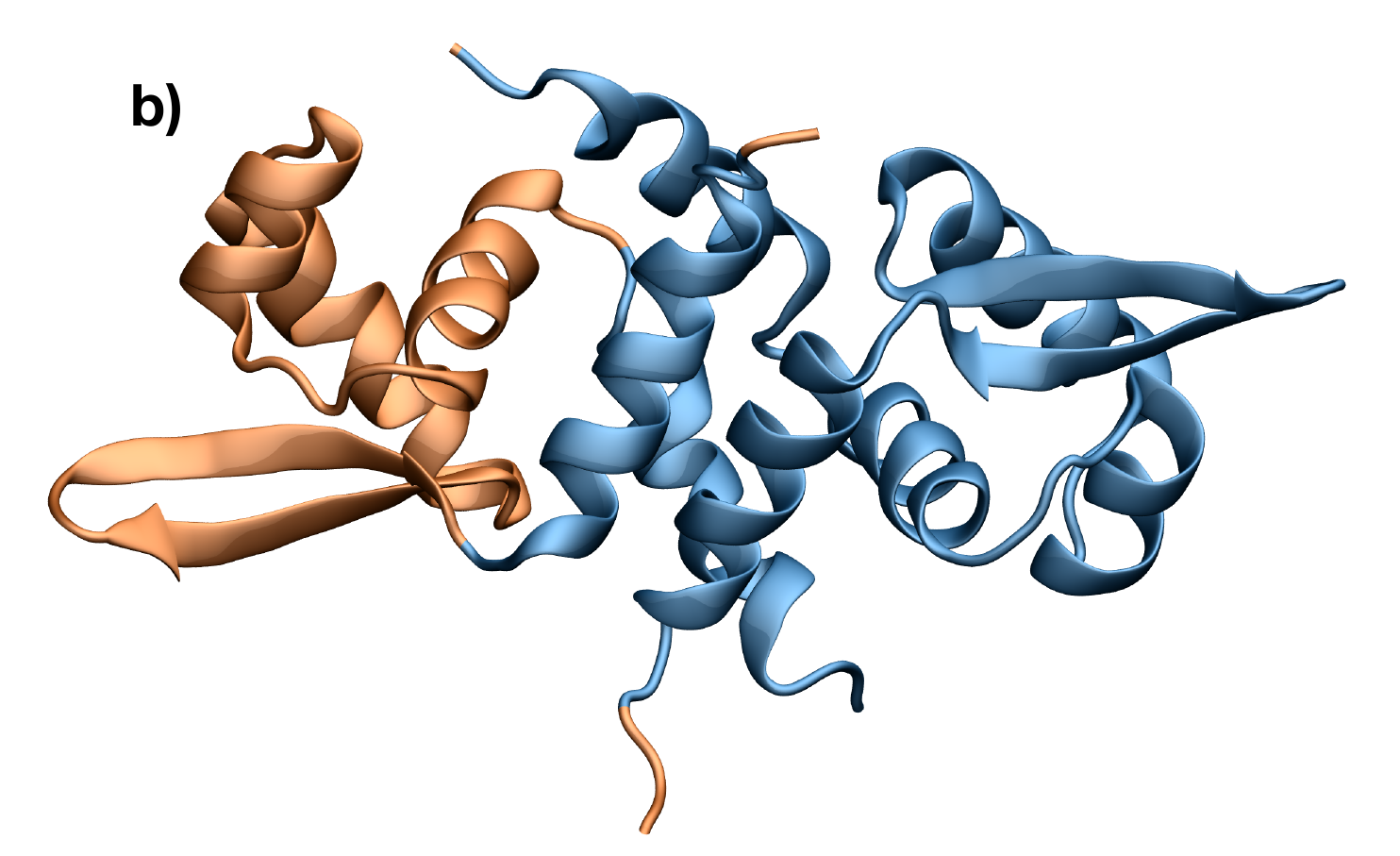}\\
\includegraphics[width=\columnwidth]{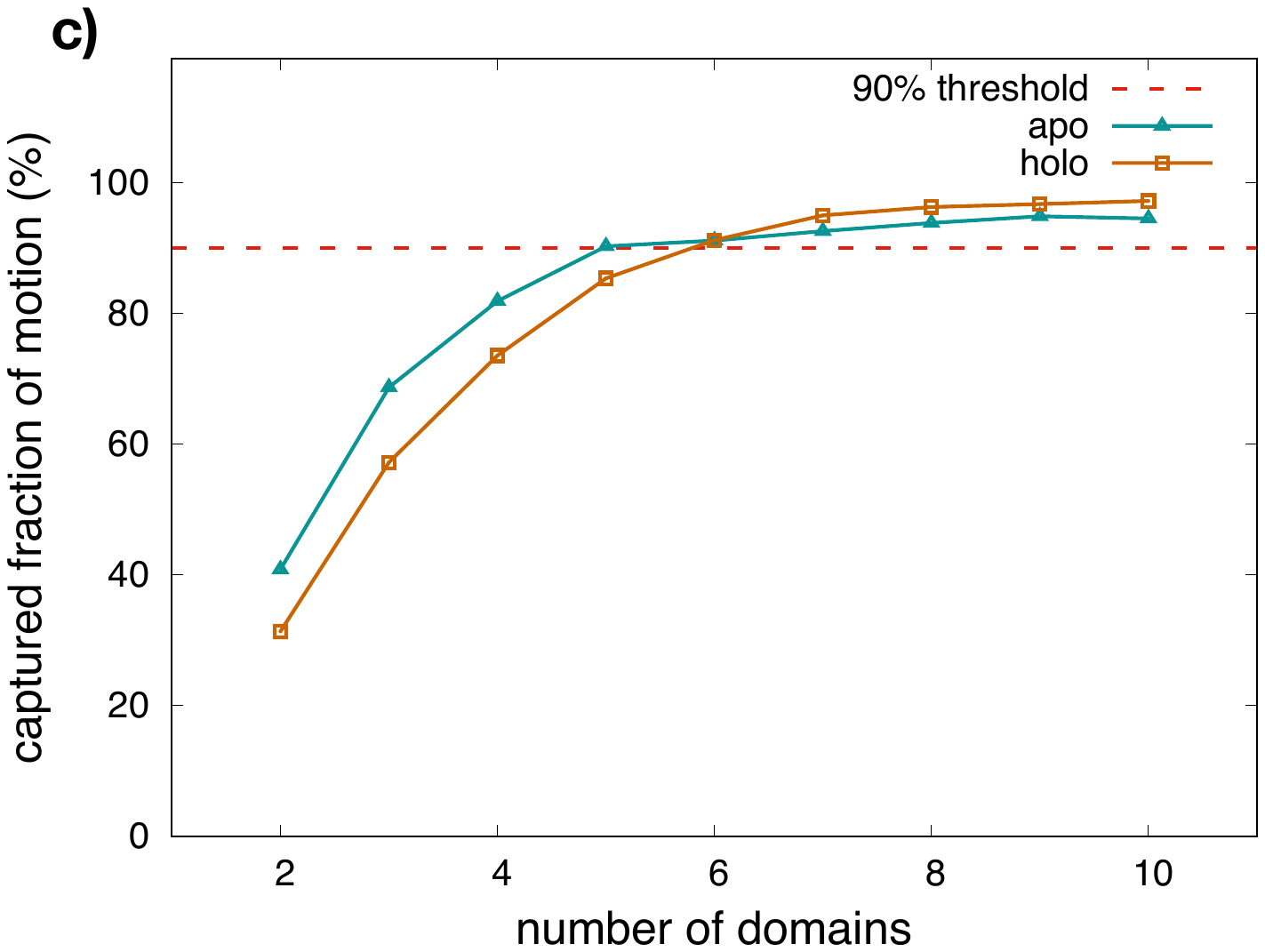}
\caption{a) Quasi-rigid domain decomposition of apo-CzrA in $2$ blocks; b) same as panel a) but in the case of the holo form; c) fraction of internal dynamics captured by inter-domain motion as a function of the number of quasi-rigid domains, for both the apo and holo systems.}
\label{fig:domains}
\end{figure}

This picture is further supported by a coarse but informative analysis of the equilibrated apo and holo structures of CzrA, obtained by relying on the protein structure quasi-rigid domain decomposition (PiSQRD) \cite{BJ2009_Potestio,BIOINFO2009_aleksiev,PLOSCB2012_morra,PLOSCB2013_polles} method. Specifically, we performed optimal partitions of the two structures in a range between $2$ and $10$ domains, where residues are assigned to a given domain so as to maximise the amount of collective dynamics that can be ascribed to inter-domain motion rather than intra-domain strain; that is to say, in an optimal partition the largest possible fraction of the internal dynamics is due to displacements of the domains \textit{relative to each other}, while the distortion \textit{within the domains} is minimised.

In Fig.~\ref{fig:domains} we compare the subdivisions of the apo (panel a) and holo (panel b) equilibrated conformations of CzrA in $2$ domains. One can easily notice the large degree of consistency between them, as well as the fact that in both cases the partition is skewed with respect to the symmetry plane of the protein as a consequence of the aforementioned asymmetric arrangement of the two monomers. Despite the similar architectures and partitions, however, the internal dynamics of the two structures differs. Indeed, the apo state features larger flexibility than the holo one, as it can be noticed from the plot in Fig.~\ref{fig:domains}c: here, we report the fraction of essential dynamics that the optimal partition entails---with a higher fraction of captured motion implying a more collective dynamics. Fig.~\ref{fig:domains}c displays that, as we subdivide the protein into an increasing number of quasi-rigid domains, the apo form remains more flexible and its dynamics more collective (and hence more easily partitioned) than the holo one up to $5$ domains, despite the two structures being characterised by similar decompositions also in these cases (data not shown). Beyond $5$ blocks, the curve for the apo state almost flattens; in contrast, the one for the holo state features a slower growth only after $7$ domains, indicating that the most relevant collective motions of the zinc-bound system take place below this structural resolution scale \cite{BJ2009_Potestio}.

These observations suggest that the coordination with the metal ions does not change appreciably \textit{how} the protein moves, but rather \textit{how much}: in fact, while the dynamics-based partition of the structure is essentially the same in the apo and holo forms, in the latter case we observe a modulation of the degree of internal flexibility of the molecule, which becomes stiffer. This result is based on a very simple description of the protein---represented in terms of a coarse-grained elastic network model \cite{micheletti2004accurate}---and its equilibrium dynamics; yet, the differences exhibited by the two states are consistent with the observed behaviour of this molecule, which, upon zinc binding from the apo state, does not manifest major conformational rearrangements but rather changes its internal motions substantially \cite{capdevila,baksh}.

The conclusion we can draw from these first analyses is that the apo and holo forms of CzrA entail noticeable differences, albeit subtle ones. An accurate assessment can highlight them, as is the case for the equilibrium dynamical features that are appreciably different in the two cases. In the following, we report a novel, complementary manner of investigating CzrA, which proved capable of bringing to light further interesting details about this system.

\subsection{Mapping entropy optimisation workflow}
\label{sec:results_mapent}

Central to the analysis method employed in the following is the concept of mapping entropy $S_{map}$ \cite{Shell2008,rudzinski2011coarse,foley2015impact,giulini,PRE2022_holtzman,giulini2021system,noid2023perspective,giulini2024excogito,aldrigo2024low}; this is a measure of the statistical information that is lost when the elements of a dataset are described in terms of a subset of their original features, see Sec.~\ref{methods:mapping_ent} for a summary of the associated theoretical details. In the context of this work, the dataset consists of the configurations sampled during computer simulations of the protein of interest; these configurations are observed through a low-resolution representation that only considers the positions of a \emph{subset} of $N<n$ atoms out of the $n$ ones composing the molecule, in what is called a \emph{decimation mapping}. By discarding constituent units from the analysis---or, more precisely, \emph{integrating them out} \cite{giulini}---one ignores part of the properties characterising the whole structure; the mapping entropy estimates the ``quality'' of such reduced representation, in that it quantifies the statistical information that a coarse description of the protein lacks with respect to the all-atom one. Critically, a specific choice of the decimation mapping one employs to inspect the system is associated with a single value of $S_{map}$, while the opposite does not necessarily hold.

From the preceding discussion, it follows that the mapping entropy is a function of the selection $\sigma_i$, $i=1,...,n$ of atoms employed to inspect the molecule, see Sec.~\ref{methods:mapping_ent}, where $\sigma_i = 1$ if atom $i$ is \emph{maintained} in the system's description and $0$ otherwise, with $\sum_{i=1}^n \sigma_i=N$ (note that we exclude hydrogen atoms from the pool of ``eligible'' ones). One can thus \emph{minimise} $S_{map}$ in the space of atom selections \cite{giulini,PRE2022_holtzman,giulini2021system,giulini2024excogito,aldrigo2024low}: this procedure aims at detecting the so-called \emph{maximally informative} reduced representations of the system, namely those whose overall information content is as close as possible to the one characterising the all-atom reference, despite being a low-dimensional projection of the latter. Given the typically rugged profile of the mapping entropy \cite{giulini}, several optimisation runs should be performed so as to collect a number $M_{opt}$ of maximally informative representations $\bar{\sigma}_i^{(k)},\ k = 1, \cdots\ M_{opt}$. From these, the frequentist probability $p_i$ for an atom to be retained in an optimal mapping can be obtained as
\begin{eqnarray}
\label{eq:average}
p_i = \frac{1}{M_{opt}} \sum_{k=1}^{M_{opt}} \bar{\sigma}_i^{(k)},
\end{eqnarray}
resulting in a probability value defined on each atom of the molecule; in the following, we will refer to this set of probabilities $p_i$ as the \textit{information field} associated with the structure. This protocol, which processes \emph{in silico} simulations of the system of interest to identify the optimal reduced representations of the latter and extract the associated information field, is dubbed mapping entropy optimisation workflow (MEOW), and is implemented for public use in the recently released EXCOGITO software suite presented in Ref.~\cite{giulini2024excogito}. A more detailed description of the approach is given in Sec.~\ref{methods:mapping_ent}.

It has been shown that the minimisation of the mapping entropy selects subgroups of atoms whose configuration is a maximally informative proxy for the global state of the whole molecule \cite{giulini,PRE2022_holtzman}. Moreover, it was observed that the atoms that are most frequently included in optimal mappings---that is, those with high probabilities $p_i$---are often highly mobile and/or display a strong energetic variability \cite{giulini}; this is rationalised by the fact that a subset of elements is informative about the rest of the system if the configurational state of the selected subgroup strongly correlates with the discarded part. This condition is more easily verified if the retained atoms entail a large conformational variability (albeit not necessarily \textit{large-amplitude} fluctuations) and interact with the others so strongly that their configuration ``dictates'' that of the remainder of the molecule \cite{aldrigo2024low}. Notably, this does not imply that all atoms that are \emph{individually} mobile and strongly interacting are selected as important: in fact, several such atoms can be ignored in optimal mappings, while few others of them are retained. This is because the kind of information that can be extracted from the MEOW analysis \textit{is an intrinsically multi-body feature of the system}: indeed, while the probability of a given atom to be retained is, by definition, an atom-specific property, it is derived from the processing of multiple optimal atom \emph{subsets}. The atoms in these subsets are identified collectively, see Sec.~\ref{methods:mapping_ent}, and the participation of an atom in a given optimal mapping cannot be deduced from properties of this atom alone (such as, e.g., type, charge, mobility, interaction strength...). In this sense, the probability field $p_i$ only represents a useful and intelligible one-body representation of an otherwise complex, intrinsically multi-body property.

Importantly, the atoms belonging to the optimal group happen to bear a nontrivial significance for the function of the molecule, as previous works have shown that the group of atoms with higher $p_i$ correlates with the biological relevance of the amino acids they belong to \cite{giulini,PRE2022_holtzman,giulini2024excogito}. It is consequently possible to employ the mapping entropy optimisation workflow with the aim of highlighting and pinpointing atoms and residues that play a critical mechanical, energetic, and functional role in a protein.

In this work, we leverage these features of the MEOW approach to investigate how the properties of the CzrA transcription repressor are influenced by the state of the zinc coordination site. Specifically, we study the system in terms of the information fields derived from the optimal low-resolution representations of its apo and holo structures (see Eq.~\ref{eq:average}), and inspect if and how the residues' retainment probabilities---and hence their functional relevance as predicted by MEOW---modulate upon binding with the metal ions. Building on and expanding the scope of previous applications of the method that only focussed on isolated, unperturbed biomolecules  \cite{giulini,PRE2022_holtzman,giulini2024excogito}, here for the first time we apply MEOW to investigate how environmental changes that take place in a system can result in a shift throughout the molecular structure of what are its biologically relevant regions.

\subsection{Application of MEOW to CzrA}

To analyse CzrA through the lenses of its maximally informative reduced representations, the MEOW pipeline of EXCOGITO~\cite{giulini2024excogito} was separately applied to both the apo and holo form MD simulations results; for each structure, a total of $10^4$ all-atom MD configurations sampled regularly from the simulation trajectories were considered in the calculations, and the number $N$ of retained sites employed to describe the system at low resolution was chosen to be equal to the number of its $C_\alpha$ atoms---that is, $187$ and $190$ for the apo and holo states, respectively. We performed $M_{opt}=48$ independent $S_{map}$ optimisations for each form of CzrA, resulting in $48$ maximally informative mappings $\bar{\sigma}_i^{(k)}$ from which the probabilities $p_i$ of each atom to be retained separately in the apo or holo systems---and hence their associated information fields---were calculated \emph{via} Eq.~\ref{eq:average}.

\begin{figure}
\centering
\includegraphics[width=\columnwidth]{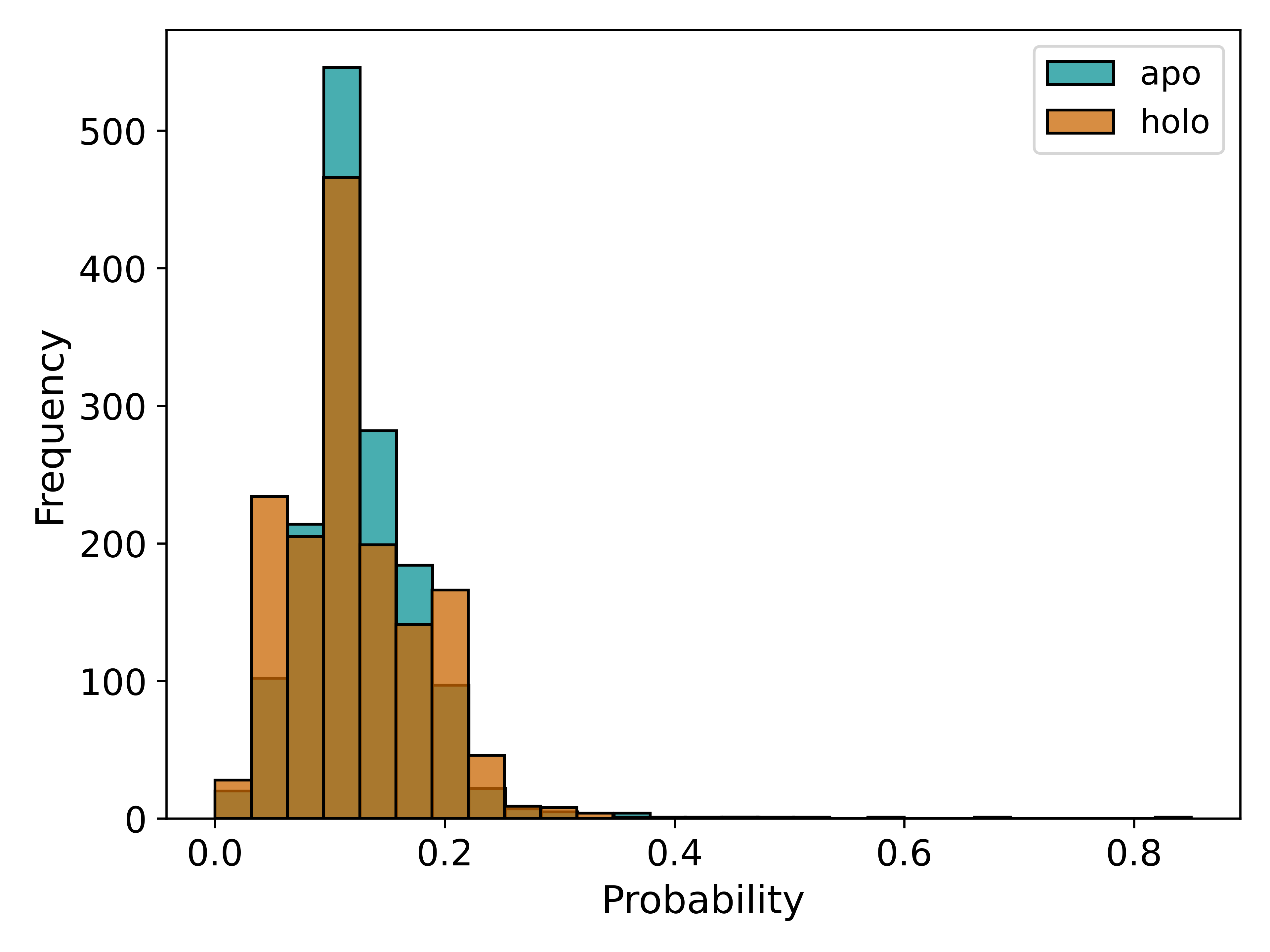}
\caption{Histogram of the values of single atom retainment probability, see Eq.~\ref{eq:average}, calculated from the pool of maximally informative representations predicted by MEOW for the apo (light blue) and holo (orange) forms of CzrA.}
\label{fig:prob_distribution}
\end{figure}

\begin{figure}
\centering
\includegraphics[width=\columnwidth]{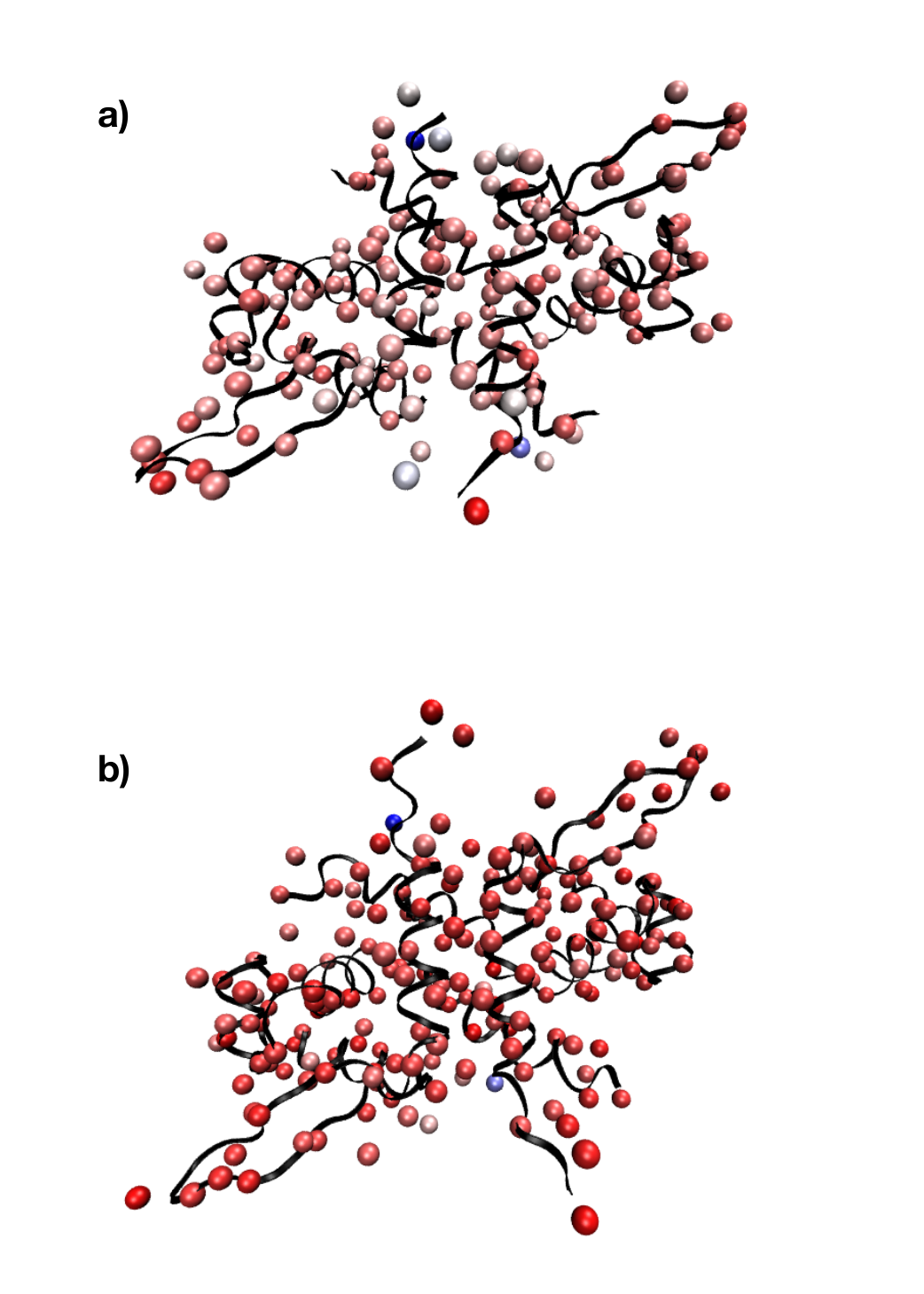}
\caption{Structural rendering of the retainment probability values of the residues of CzrA (the MEOW \emph{information field}, see main text). For each amino acid, a bead is located in the position of the atom that displays the highest probability and is coloured according to the latter. In red, low probability of being retained in an optimal mapping; in white, intermediate values; in blue, high probability. a) Results for the apo form; b) results for the holo form.}
\label{fig:heatmap_structure}
\end{figure}

The histograms of the sets of $p_i$ obtained from the MEOW analysis, shown in Fig.~\ref{fig:prob_distribution}, highlight that the apo and holo forms of CzrA are roughly compatible in terms of \emph{how many} (heavy) atoms appear
with a given value of $p_i$ in the pool of optimised reduced representations of the protein. In both cases, the distribution is approximately unimodal, with the most prominent peak being centred around $p\sim 0.13$. Atoms located to the left of such peak are predominantly \emph{excluded} from the pool of optimised mappings of the corresponding form of CzrA; conversely, the relatively long tails that extend from the mode of the distributions up to high probability values pertain to atoms that, according to MEOW, need to be retained when describing the system at a coarser level of detail in order to preserve the maximum amount of information about its statistical behaviour. The next natural step in the analysis is thus to investigate how the two ensembles of single-atom probabilities $p_i$ \emph{distribute} throughout the molecular structure, and if dissimilarities in the MEOW information fields that result from these projections are present due to zinc coordination. To this aim, in Fig.~\ref{fig:heatmap_structure} we separately show the apo and holo forms of CzrA rendered as thin ribbons decorated with beads; each bead is representative of a single amino acid and is located on the position of the \emph{most frequently conserved} heavy atom in the residue. Beads are further coloured according to the value of their underlying atom's $p_i$, going from red to blue (through white) in transitioning from low to high probabilities. 

An inspection of Fig.~\ref{fig:heatmap_structure} reveals that, in both states, the projection results in a fairly smooth colouring pattern throughout the protein structure, hence with whole regions of the system that are identified by MEOW as being more or less informative. Significant differences are however appreciable between the probability fields associated with the apo and holo states of the molecule. Most notably, we previously discussed how the overall \emph{amount} of atoms displaying a specific value of retainment probability is approximately compatible in the two protein forms, see Fig.~\ref{fig:prob_distribution}; critically, Fig.~\ref{fig:heatmap_structure} reveals that, in contrast to the apo case, upon zinc coordination a large number of \emph{residues} in the protein end up being characterised by a relatively low $p_i$. As only the \emph{most likely conserved} heavy atom of each amino acid has been explicitly depicted in Fig.~\ref{fig:heatmap_structure}, this result suggests that the information content of the holo form of CzrA is somewhat concentrated in more localised regions of the molecular structure compared to its apo counterpart. We remind that the analyses performed in Sec.~\ref{sec:structural_analysis} highlighted that the binding with the metal ions induces a significant reduction in the mobility of the protein around its average conformation, see Fig.~\ref{fig:rmsf}. Interestingly, MEOW hints that such stiffening goes on par with a remodulation of the system's \emph{energetic frustration}; starting from a quite spread distribution of the latter throughout the molecular structure in the apo case (resulting in rather diffused peaks in the MEOW information field), upon zinc coordination localised regions characterised by a high energetic variability instead appear, which stand out of what is otherwise a rather ``silent'' background and acquire high relevance as prescribed by the mapping entropy approach. To gain deeper insight into the effects of ion binding onto CzrA, let us now further analyse the discrepancies in the information fields of the two protein forms; in particular, we focus our attention on the regions of the system with a known biological function, namely those involved in the coordination with the metal and the binding with DNA.

\begin{figure}
\centering
\includegraphics[width=\columnwidth]{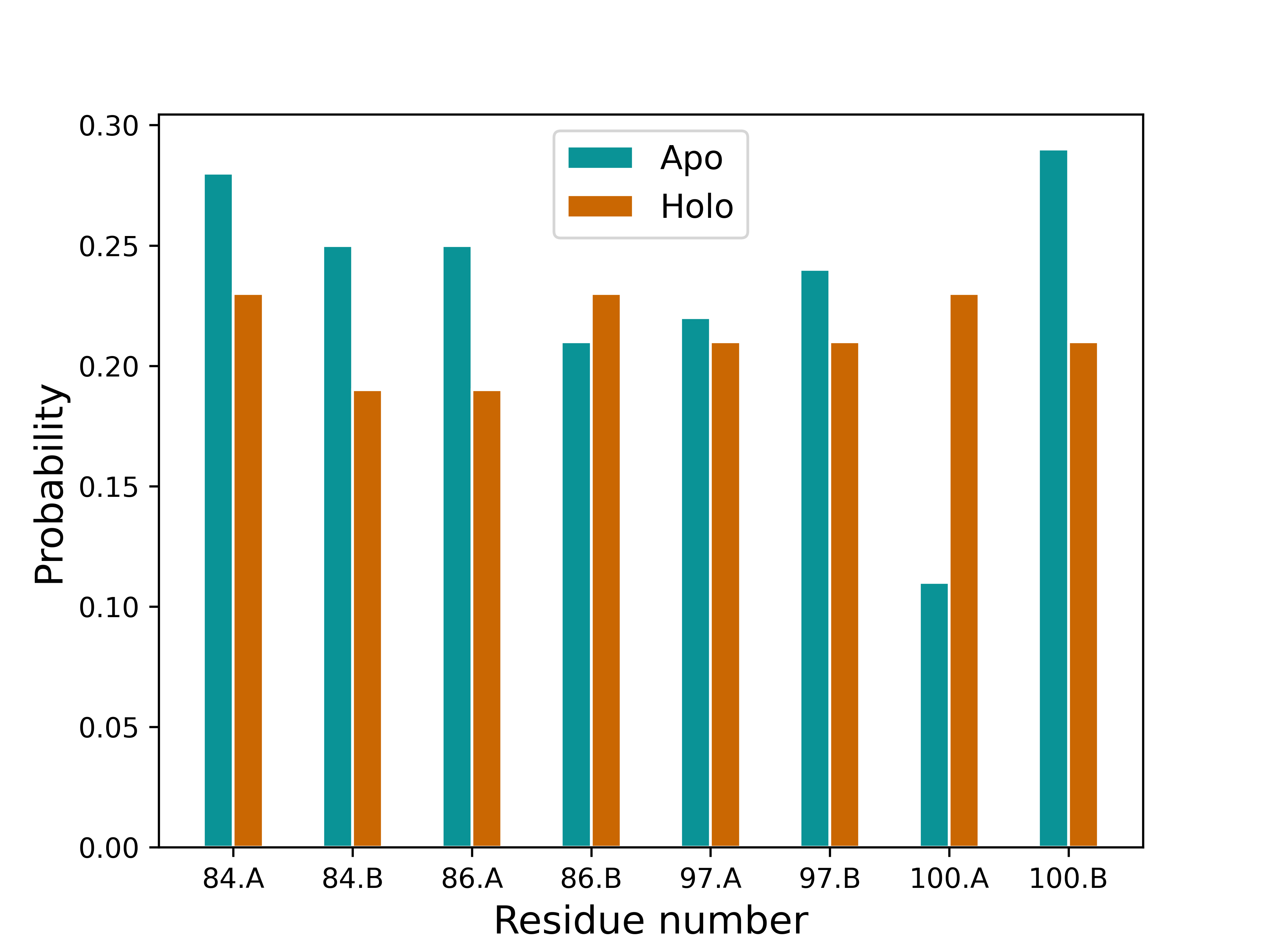}
\caption{MEOW predictions of the probability of retaining a residue for the amino acids of CzrA involved in zinc coordination. The probability associated with each residue is the one of its most conserved heavy atom. We display results for the two chains of the molecule (\#.A and \#.B, \# being the residue number in each chain) in both the apo (light blue) and holo (orange) forms.}
\label{fig:prob_zn_sites}
\end{figure}

The conservation probabilities of the residues belonging to the zinc coordination site of CzrA are displayed in Fig.~\ref{fig:prob_zn_sites} separately for the apo and holo states; we recall that this per-residue probability is defined, in each system, as the largest $p_i$ value among the (heavy) atoms belonging to the selected amino acid. Despite the presence of slight differences between the results of the two protein chains due to the structural asymmetry discussed in Sec.~\ref{sec:structural_analysis}, we observe that the $p_i$ of the zinc coordination residues are generally higher than the mode of the histograms reported in Fig.~\ref{fig:prob_distribution}; most notably, almost all probability values in the apo case are larger than $0.2$, hence being located in the right tail of the associated distribution. This indicates that these residues entail a large information content about the system, a remarkable fact in that, while their functional role as zinc coordination regions is known empirically, the MEOW protocol unveils their importance only starting from raw MD simulations data in an unsupervised manner.

Moreover, we observe that the majority of the binding site residues are retained more frequently in the apo state of the molecule than in the holo one, with MEOW thus highlighting a decrease in their relevance upon coordination with the metal. Such a decrease can be ascribed to the strong interaction of these residues with the zinc ions, which reduces their original mobility (see the RMSF in Fig.~\ref{fig:rmsf}) as well as energetic frustration, thus negatively impacting two pivotal features that in the apo state drove the protocol to preserve these regions in order to minimise the mapping entropy, see Secs.~\ref{sec:results_mapent} and~\ref{methods:mapping_ent}. Consequently, in the holo form these amino acids return a lesser amount of information about the behaviour of the protein as quantified by $S_{map}$.

\begin{figure}
\centering
\includegraphics[width=\columnwidth]{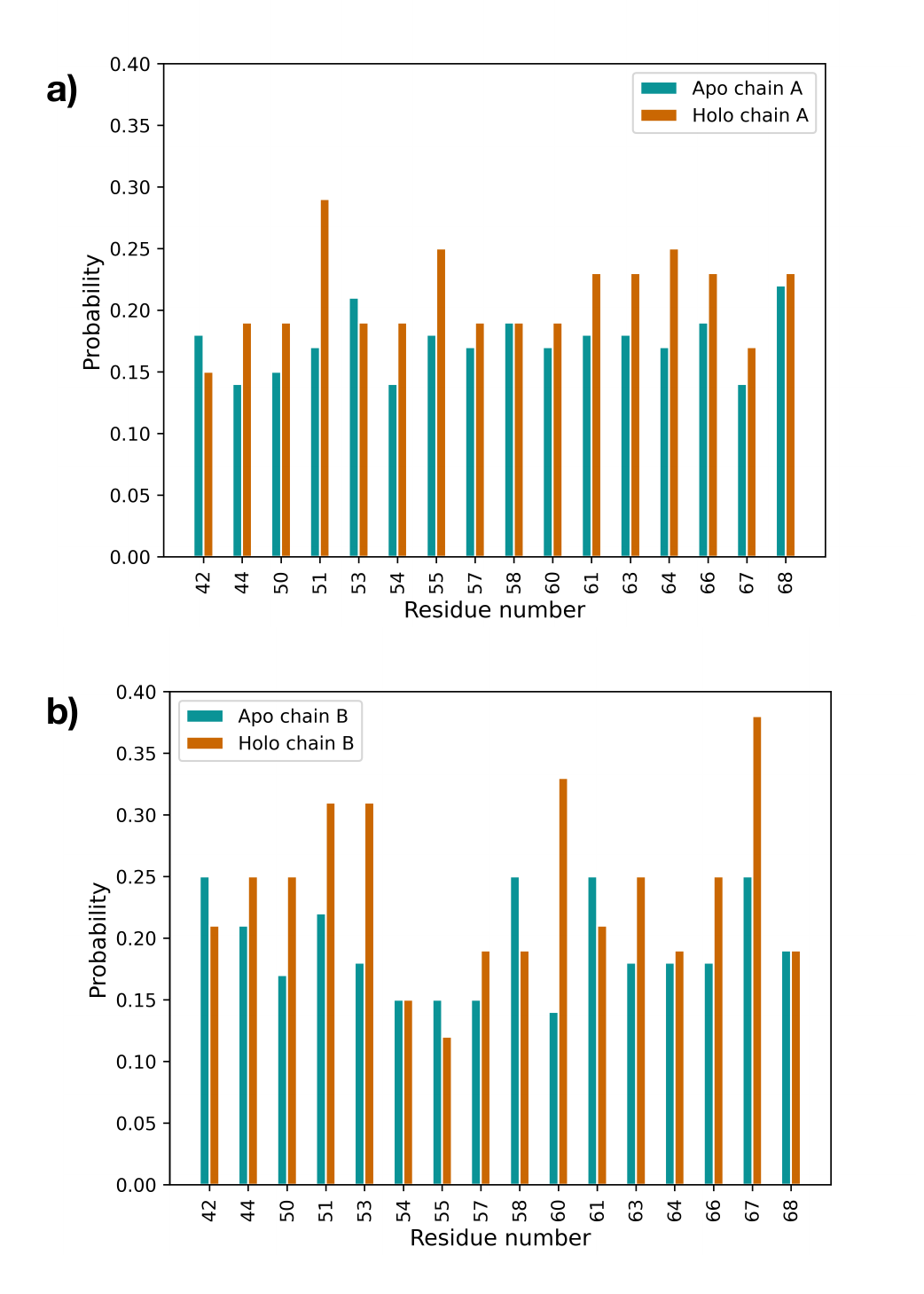}
\caption{MEOW predictions of the probability of retaining a residue for the amino acids of CzrA involved in DNA binding. The probability associated with each residue is the one of its most conserved heavy atom. We display results for both the apo (light blue) and holo (orange) forms of the protein, separately for chain A in panel a) and chain B in panel b).}
\label{fig:prob_dna_sites}
\end{figure}

Even more compelling is the MEOW analysis of the DNA binding region of CzrA, for which, opposite to what was observed in the case of the zinc coordination site, the results highlight an \emph{increment} in the relevance of the associated residues when the molecule is bound to the metal ions. The MEOW information fields of the DNA binding region in the two structures, summarised in Fig.~\ref{fig:prob_dna_sites}, indeed show that its amino acids are generally \emph{more frequently} conserved when CzrA is in complexation with the zinc ions than in the absence of the latter. Also in this case differences are present between the details of the probability patterns of the two protein chains due to their structural asymmetry; this notwithstanding, the overall increase of the information field in going from the apo to the holo state can be appreciated in both units. Notably, while the $p_i$ values for the apo state of the DNA binding region are on average slightly below the corresponding ones of the zinc coordination site (see Fig.~\ref{fig:prob_zn_sites}), their boost in the holo state is such that they reach peaks as high as $p_i \simeq 0.30-0.37$, hence being located in the far right of the tail of the holo distribution reported in Fig.~\ref{fig:prob_distribution}. This increment in information field values upon zinc coordination takes place in the presence of an appreciable reduction in the amplitude of the DNA binding region's fluctuations \cite{arunkumar2009solution,Chakravorty}, see Fig.~\ref{fig:rmsf}. As previously discussed, the minimisation of $S_{map}$ highlights those residues such that, once their configuration is fixed, the structural and energetic variability of the remainder of the molecule is constrained as much as possible. Hence, an increase of the $p_i$ of the residues in the DNA binding region in the holo state is indicative of the fact that, in spite of a reduced amplitude of their mobility, they acquire greater energetic frustration, and hence greater MEOW relevance, in what is a more conformationally restrained state of the molecule.

In summary, the application of the MEOW approach to the configurations sampled in the MD simulations of the apo and holo forms of CzrA has shown important differences between them, which can elude a point-wise analysis of the structural and dynamical features of the corresponding residues.

\section{Conclusions}
\label{sec:conclusions}

Allosteric regulation is one of the most relevant transduction mechanisms at the molecular level \cite{Chemrev2016_nussinov,PLOSCB2016_liu,WODAK2019566}. The system-wide modifications that one or few atoms, interacting in very local regions, induce on the whole protein make allostery an exquisitely multi-scale process, whose versatility and consequent ubiquity are paired by the complexity of its investigation. Ion binding in metal sensor proteins is a brilliant example of the exploitation of this multi-scale strategy, in which the affinity of the molecule to the DNA substrate is modulated through the control of the distal ion coordination site \cite{giedroc2007metal,reyes2011metalloregulatory}.

In this work, we have addressed the study of the ArsR/SmtB transcription repressor factor CzrA. This is a metal-sensor protein, in which the coordination with zinc ions determines large-scale alterations of the system's structural and dynamic properties, further leading to a state of low affinity towards the DNA substrate \cite{arunkumar2009solution,Chakravorty}. Specifically, we focused on the differences between the apo and holo (zinc-bound) states of the protein in the absence of the DNA, analysing $\mu$s-long all-atom molecular dynamics simulations of these two systems through well-established techniques as well as more recent approaches.

First, we studied the consequence of ion binding on the structural arrangement of the protein as well as on its large-scale flexibility. Our analyses have shown virtually no discrepancies between the protein's overall architecture in the presence or absence of the zinc ions; indeed, the average conformation of the molecule in the apo and holo states differ by as little as $0.11$ nm. 
This notwithstanding, the occasional excursions to more ``closed'' configurations that the protein was observed to undergo in the apo form, as well as the seldom transitions to more ``open'' states that appeared in the presence of the ions, suggests a tendency of the holo state to display a reduced affinity towards the DNA substrate compared to its apo counterpart. Further inspection of the trajectories has highlighted subtle yet marked changes in the equilibrium dynamics of the protein that occur upon coordination with the metal; these are quite evident in the local conformational variability of the molecule, and most prominently consist in a sensible reduction of its overall flexibility when in the zinc-bound state.

These results are consistent with, and corroborate previous computational \cite{Chakravorty} as well as experimental works \cite{capdevila,baksh}; in particular, in the former, through the analysis of MD simulations of the apo and holo systems it was shown that the most prominent differences between these two states of the protein, rather than in their structural properties, are to be observed in their dynamics, with the DNA-bound state being the only one in which a major conformational rearrangement of the molecule occurs. Our observations support this picture through the analysis of more extensive simulations.

Additionally, it was suggested in various experimental works \cite{capdevila} that the allosteric mechanism employed by CzrA does not involve important distortions in the overall organization and architecture of the molecule, but rather relies on a diffused redistribution of the residue-specific conformational entropy of the system in going from the apo to the holo state. This hypothesis found further consistency in the analysis we carried out in this work by relying on the recently-developed mapping entropy optimisation workflow \cite{giulini,PRE2022_holtzman,giulini2021system,giulini2024excogito}. MEOW analyzes an MD trajectory of a protein to identify optimal low-resolution representations, or optimal mappings, of the system; these are particular subsets of the molecule's atoms, such that the amount of information about the conformational space they sample is as close as possible to the one associated to the full protein.
From the atom-wise probability of being part of an optimal mapping, one can construct what we refer to as the protein's \emph{information field}, which assigns to each atom a measure of its relevance in what is a synthetic yet informative description of the molecule. Most importantly, the information field returned by the MEOW protocol was previously found to single out regions of particular structural, energetic, and functional relevance of a system \cite{giulini,PRE2022_holtzman,giulini2024excogito}; in this work, we have for the first time exploited this feature to investigate how such relevance is affected by changes in the system's environmental conditions. 

The MEOW analysis of CzrA has shown that, by coordinating with the ion substrates, the information pattern of the molecule redistributes throughout the structure, and concentrates in rather localized areas that are hence characterised by an increased energy frustration. Such redistribution is particularly interesting when one looks in detail at the regions of the molecule playing a key biological role, namely its binding sites; focusing on the zinc coordination site, we found that the informativeness associated with its residues \emph{decreases} when going from the apo to the holo state. This change is suggestive of the fact that the presence of the ions somehow ``deactivates'' the zinc coordination site. At the same time, the MEOW approach highlighted an opposite behaviour in the residues of the DNA binding region depending on the state of the distal zinc coordination site. Specifically, we have observed that the level of informativeness of the former \emph{is larger} in the holo conformation than in the apo one, thus implying their ``activation'' in the presence of the metal ions---but in absence of the DNA substrate.

As anticipated, the remodulation of the information field of the molecule upon zinc binding can be interpreted in terms of an overall change in its structural and energetic frustration. In fact, we observed that the presence of the zinc ions determines an overall increase in the stiffness of the protein, which affects the binding sites of both substrates making them less mobile; this notwithstanding, the informativeness of the DNA binding region increases in the holo form, thereby signaling a greater conformational and energetic variability in spite of a reduced \emph{amplitude} of its fluctuations. This variability is instrumental in the operation of the allosteric mechanism that allows the protein to release from the DNA filament; notably, such a picture is consistent with the previously observed entropy remodulation that the molecule undergoes upon zinc binding \cite{capdevila,baksh}.

We stress here that the MEOW protocol has highlighted the change in the properties of the residues involved in the interaction with the DNA \emph{in the absence of this substrate}, thereby providing novel and complementary insight into the allosteric mechanism employed by CzrA. This makes the MEOW approach a promising candidate to complement and expand the scope of the already available methods for the study of binding sites in proteins \cite{ebert2008robust, hu2016recognizing, lu2022mib2}.

In conclusion, these results contribute interesting information about the behaviour of CzrA, and demonstrate, in addition to previous applications \cite{giulini,giulini2024excogito}, the validity of the MEOW analysis pipeline to characterize the properties of key residues in a protein and rationalise its global behaviour also through the comparison of the information fields across different states of the molecule. We thus foresee fruitful applications of the proposed approach to gain insight into the way proteins perform and modulate their biological function.

\section*{Acknowledgments}

The authors are indebted to Marco Giulini for technical support and an insightful reading of the manuscript.
RP acknowledges support from ICSC - Centro Nazionale di Ricerca in HPC, Big Data and Quantum Computing, funded by the European Union under NextGenerationEU. Views and opinions expressed are however those of the author(s) only and do not necessarily reflect those of the European Union or The European Research Executive Agency. Neither the European Union nor the granting authority can be held responsible for them.
Funded by the European Union under NextGenerationEU. PRIN 2022 PNRR Prot. n. P2022MTB7E.

\section*{Data and software availability}

Raw data produced and analyzed in this work are freely available on the Zenodo repository \href{https://zenodo.org/records/10700290}{https://zenodo.org/records/10700290}.

\section*{Author contributions}

MR and RM proposed the study; RP and RM conceived the work plan and proposed the method; MR carried out the simulations; MR and RM carried out the preliminary data analyses. All authors contributed to the analysis and interpretation of the data. All authors drafted the paper, reviewed the results, and approved the final version of the manuscript.

\section{Methods}
\label{sec:methods}

\subsection{Protein systems setup and simulations}
\label{sec:methods_allatom}

In this work, MD simulations of CzrA were performed starting from the experimentally resolved structures of its apo and holo forms. The 3d configuration of the apo system (PDB code 1R1U) was determined through X-ray diffraction with a resolution of $2$~\AA~\cite{Eicken}, while the structure of the holo state (PDB code 2M30) consists of a complex formed by the CzrA protein and two Zn$^{2+}$ ions, and was determined \emph{via} NMR spectroscopy refined through quantum mechanics/molecular mechanics (QM/MM) simulations \cite{Chakravorty1}. \emph{In silico}, we individually solvated the two systems in water using the TIP3P model \cite{tip3p}, adding Na and Cl ions to neutralise the total charge and to mimic the physiological salt concentration ($150$~mM). The simulation box for the apo and holo forms was chosen of dodecahedron shape, with the protein having a minimum distance of $1.1$~nm from the box edge in both cases. After energy minimisation and proper equilibration, the production runs of each system were then performed in the NPT ensemble at $300$~K and $1$~bar through the stochastic velocity-rescale thermostat \cite{bussi1} and the Parrinello-Rahman barostat \cite{parrinello_bar}, respectively with a temperature coupling constant $\tau_t=0.1$~ps and a pressure coupling constant $\tau_p = 2$~ps. The integration time step was set to $2$~fs, the selected integrator was leap-frog, and holonomic constraints were accounted for by means of the LINCS algorithm \cite{lincs}. The two forms of CzrA were simulated for $1$~$\mu$s of production run each by relying on the  Amber14sb \cite{amber14} force field. Finally, we included position restraints in the setup of the holo system to mimic the binding of the two zinc ions to their respective coordination sites and maintain them in the correct position. The atoms involved in the interaction with the first Zn$^{2+}$ ion are ASP 84.B OD2, HIS 86.B ND1, HIS 97.A ND1, HIS 100.A NE2. As for the second zinc atom, the restraints were put between the ion and ASP 84.A OD2, HIS 86.A ND1, HIS 97.B ND1, HIS 100.B NE2. The functional form of the restraining potential was 
\begin{numcases}{V(r_{ij})=}
\label{eq:Vres}
0  &$r_{ij}<r_{ij}^0, $ \nonumber \\ 
\frac{1}{2}k_{dr}(r_{ij}-r_{ij}^0)^2 &$r_{ij}^0\le r_{ij}<r^1, $  \nonumber  \\ 
\frac{1}{2}k_{dr}(r^1-r_{ij}^0)\times \nonumber \\ \times (2r_{ij}-r^1-r_{ij}^0) &$r_{ij}\ge r^1, $
\end{numcases}
where $r_{ij}$ is the distance between the two atoms involved in the interaction, $r_{ij}^0$ is the corresponding distance measured in the experimental 2M30 structure, $r^1$ is the maximum length set to $0.35$~nm, and k$_{dr}$ is set to $2 \cdot 10^3$~kJ/mol/nm$^2$.

\subsection{Mapping entropy and MEOW analysis tool}
\label{methods:mapping_ent}

The mapping entropy optimisation workflow is a tool developed by Giulini and coworkers that aims at identifying functional regions of a biomolecular system only starting from raw MD simulation data \cite{giulini,PRE2022_holtzman,giulini2021system}. The protocol relies on the concept of coarse-graining \cite{rudzinski2011coarse,noid2023perspective}, interpreted as the analysis of the biomolecule's configurational space \textit{via} the projection of the latter on a restricted subset of the original degrees of freedom. In general, such a procedure entails a loss of information on the system's statistical properties, which critically depends on the choice of the projection and is quantifiable in terms of the mapping entropy $S_{map}$ to be briefly summarised in the following \cite{Shell2008,rudzinski2011coarse,foley2015impact,giulini,PRE2022_holtzman,giulini2021system,noid2023perspective,giulini2024excogito}; the aim of MEOW is to detect the low-resolution representations for which the aforementioned information loss---and accordingly the mapping entropy---is as small as possible, and are hence maximally informative about the statistical behaviour of the system despite a coarsening of its structure.

The type of coarse-graining we rely on in this work is a \emph{decimation}, in which the low-resolution representation in terms of which the molecule is inspected is obtained by selecting a subset of $N$ atoms out of its $n$ constituent ones. This can be expressed via a set of binary variables $\boldsymbol{\sigma}=\{\sigma_i\}$, $i=1,...,n$, where $\sigma_i =1$ or $0$ depending on whether atom $i$ is maintained or neglected in the system's description, and $\sum_{i=1}^n \sigma_i=N$. A specific choice of which atoms are retained constitutes a \emph{mapping}; in the formal theory of coarse-graining, the latter is defined through a projection operator ${\bf M}$ that converts a high-resolution configuration ${\bf r}=\{{\bf r}_i\}$ of the system---where ${\bf r}_i$ are the three-dimensional Cartesian coordinates of atom $i$---into a coarser configuration, or \emph{macrostate} ${\bf R}=\{{\bf R}_I\}$, $I = 1,...,N < n$ given in terms of the fewer atoms that were selected \cite{rudzinski2011coarse,giulini2021system,noid2023perspective}, with
\begin{equation}
\label{eq:mapping_firstdef}
    {\bf M}_{I}({\bf r}) = {\bf R}_{I} = \sum_{i=1}^{n} c_{Ii}  {\bf r}_i.
\end{equation}
In the case of a decimated representation, in Eq.~\ref{eq:mapping_firstdef} $c_{Ii} = 1$ if atom $i$ is retained (and thus mapped onto the coarse-grained site $I$), and $0$ otherwise. 

Critically, the filtered version of the system one obtains through such coarse-graining procedure is characterised by an information loss on what are the statistical properties of the reference, high-resolution structure; for a specific selection of atoms employed to describe the biomolecule, this loss can be measured \emph{via} the associated mapping entropy $S_{map}$ defined as \cite{Shell2008,rudzinski2011coarse,foley2015impact,giulini,PRE2022_holtzman,giulini2021system,noid2023perspective,giulini2024excogito}
\begin{eqnarray}
\label{eq:smap_main}
S_{map}(\boldsymbol{\sigma}) = S_{map}(\mathbf{M}) = k_B\int d{\bf r}\ p_r({\bf r}) \ln \left[ \frac{p_r({\bf r})}{\bar{p}_r({\bf r})} \right].
\end{eqnarray}
We observe that the mapping entropy is a (non-negative) Kullback-Leibler divergence between two probability distributions \cite{kullback1951information}; the first, $p_r({\bf r})$, is the one characterising the original high-resolution system, and in the case of thermal equilibrium it is given by the Boltzmann measure:
\begin{eqnarray}
\label{eq:pmicro}
&& p_{r}({\bf r})=\frac{1}{Z}\ e^{-\beta u({\bf r})}, \\
&& Z =\int d{\bf r}\ e^{-\beta u({\bf r})},
\end{eqnarray}
where $\beta=1/k_BT$, $u({\bf r})$ is the microscopic potential energy of the system, and $Z$ its (configurational) canonical partition function. The probability $\bar{p}_r({\bf r})$ in Eq.~\ref{eq:smap_main}, on the other hand, reads \cite{Shell2008,rudzinski2011coarse,foley2015impact,giulini,PRE2022_holtzman,giulini2021system,noid2023perspective,giulini2024excogito}
\begin{equation}
\label{eq:omega}
\bar{p}_r({\bf r}) = \frac{p_R({\bf M}({\bf r}))}{\Omega_1({\bf M}({\bf r}))}
\end{equation}
and represents the statistical description of the all-atom system that one would obtain in an attempt to reconstruct the properties of the latter \emph{only starting from a knowledge of its low-resolution, filtered counterpart}, thus reverting the coarse-graining procedure. In Eq.~\ref{eq:omega}, $p_R({\bf R})$ is the probability to sample the CG macrostate ${\bf R}$, given by
\begin{equation}
\label{eq:pmacro1}
 p_R({\bf R})=\int d{\bf r}\ p_{r}({\bf r})\delta({\bf M}({\bf r}) - {\bf R}),
\end{equation}
while $\Omega_1({\bf R})$ is defined as
\begin{equation}
\label{eq:omega1}
\Omega_1({\bf R}) =   \int d{\bf r}\  \delta({\bf M}({\bf r}) - {\bf R}).
\end{equation}
$\Omega_1({\bf R})$ indicates the degeneracy of the macrostate, that is, the number of microstates $\bf{r}$ that map onto the same CG macrostate $\bf{R}$. Eqs.~\ref{eq:omega}-\ref{eq:omega1} clarify the origin of the loss of information generated by coarse-graining: indeed, all the microscopic configurations that enter the composition of a specific macrostate become statistically equivalent upon backmapping, with the probability $\bar{p}_r$ that is common to all of them being given by the average of their original probabilities $p_r({\bf r})$. The mapping entropy in Eq.~\ref{eq:smap_main} quantifies this loss globally \emph{via} a Kullback-Leibler divergence between the reconstructed, smeared all-atom distribution and the ``genuine'', detailed one. It is crucial to underline that $S_{map}$ depends on the selection of atoms $\boldsymbol{\sigma}$ in a multi-body fashion, simultaneously tethering together all the $N$ constituents that are employed to describe the molecule at a lower resolution. This is a consequence of the probability $p_R({\bf R})$ in Eq.~\ref{eq:pmacro1} entering the calculation of the mapping entropy: due to the coarse-graining procedure, $p_R({\bf R})$ is, in fact, intrinsically $N$-body in nature, not simply factorizable as the product of distributions of lower order even if the original all-atom system comprises, e.g., only pair interaction potentials acting among its constituents. \cite{dijkstra1999phase,likos2001effective,d2015coarse,menichetti2017thermodynamics}. 

The aim of MEOW is thus now to determine, among all the possible selections $\boldsymbol{\sigma}$ of atoms that can be designed to describe the system at a lower resolution, those that retain the largest amount of information about the all-atom reference---and hence minimise the mapping entropy. The first step in such an analysis is, quite naturally, estimating the $S_{map}$ associated with a \emph{specific} choice of the CG mapping. To fulfil this task, rather than on Eq.~\ref{eq:smap_main}, in this work we rely on the approximate expression derived by Giulini and coworkers \cite{giulini}, which enables the calculation of $S_{map}$ only provided a set of all-atom configurations sampled from $p_r({\bf r})$ \emph{via}, e.g. an MD simulation, as well as the selected CG representations. More specifically, $S_{map}$ is evaluated as a weighted average, over all CG macrostates $\mathbf{R}$, of the variance of the atomistic potential energies of all configurations $\mathbf{r}$ that map onto a specific macrostate. We refer the interested reader to Ref.~\cite{giulini} for the theoretical details and an in-depth discussion of the algorithmic implementation of the resulting $S_{map}$ estimation workflow, further reminding that the latter is included in the freely available EXCOGITO software suite described in Ref.~\cite{giulini2024excogito}.

With this ingredient at hand, we now move to the identification of the maximally informative reduced representations of the system that \emph{minimise} the mapping entropy. In principle, this could be achieved by exhaustively probing all the possible selections $\boldsymbol{\sigma}$ of subsets of $N$ atoms within the molecular structure, ranking them according to their value of $S_{map}$. The size of the CG mapping space is, however, overwhelmingly large (if one retains, e.g. $100$ atoms out of a protein of $100$ amino acids, the number of possible selections to be probed in this scheme would be $\sim 10^{272}$) preventing the tackling of the optimisation problem via simple enumeration. $S_{map}$ is thus minimised in the space of possible CG representations of the system with $N$ retained atoms through a Monte Carlo simulated annealing protocol \cite{Kirkpatrick}, see Refs.~\cite{giulini,giulini2024excogito} for all technical details. As the $S_{map}$ landscape is likely rugged and prone to have a large number of more or less degenerate local minima \cite{giulini}, several optimisations are further performed to gain a robust and informative idea about the landscape of minimum information loss for the system at hand, resulting in a \emph{pool} of $M_{opt}$ different optimal mappings $\bar{\boldsymbol{\sigma}}^{(k)},\ k = 1, \cdots\ M_{opt}$. These are employed as described in Eq.~\ref{eq:average} to compute the frequentistic probability with which a given atom is found in a CG representation of minimal mapping entropy---the so-called \emph{information field}. As it was shown in various works \cite{giulini,PRE2022_holtzman,giulini2024excogito}, this protocol is found to be able to provide an ensemble of low-resolution descriptions of the system retaining atoms that are important from the biological point of view, e.g. atoms involved in biochemical functions such as substrate binding or catalysis.

\bibliography{main.bib}

\end{document}